\documentclass[prd,preprint,superscriptaddress,amsmath,amssymb,nofootinbib]{revtex4}
\usepackage{graphicx}
\usepackage{dcolumn}
\usepackage{bm}
\usepackage{amssymb}
\usepackage{amsmath}
\usepackage{epsfig}    
\usepackage{color}
\usepackage{slashed}
\usepackage{hhline}
\usepackage{bbm}
\usepackage{tikz-feynman}
\tikzfeynmanset{compat=1.1.0}

\def\be{\begin{equation}}
\def\ee{\end{equation}}
\newcommand{\bea}{\begin{eqnarray}}
\newcommand{\eea}{\end{eqnarray}}
\newcommand{\nn}{\nonumber}

\begin{document}

\title{Dynamical Determination of the Cut-off Scale in Loop-Induced Neutrino Mass Models with Non-Invertible Symmetry}

\author{Hiroshi Okada}
\email{hiroshi3okada@htu.edu.cn}
\affiliation{Department of Physics, Henan Normal University, Xinxiang 453007, China}

\author{Jia-Jun Wu}
\email{wujiajun@htu.edu.cn}
\affiliation{Department of Physics, Henan Normal University, Xinxiang 453007, China}

\date{\today}

\begin{abstract}
 We propose {our framework as an effective field theory valid below the cut‑off scale 
 $\Lambda$, in which }
we explain the tiny scale of neutrino masses by integrating a non-invertible symmetry 
 with the dynamical determination of the cut-off scale. In our model, we introduce three families of SU(2)$_L$ quintet fermions ($\Sigma_R$) and a quartet scalar ($\phi_4$), both of which are charged under the Fibonacci fusion rule (FFR).
 A central feature of this construction is that the vacuum expectation value (VEV) $v_4$ of $\phi_4$ is induced at the one-loop level via dynamical symmetry breaking. To resolve the inherent arbitrariness of the cut-off scale $\Lambda$ in loop-induced VEV models, we identify $\Lambda$ with the scale at which the SU(2)$_L$ gauge coupling $g_2$ encounters the renormalization group evolution (RGE) of $g_2$, naturally fixing the physical cut-off within the range of approximately $10^5$ to $10^7$ GeV.
 Quantitatively, this framework yields 
 $v_4\sim0.07-0.1$  GeV, which in turn leads to neutrino Yukawa couplings on the order of $10^{-3}$.
 This result provides a significantly more natural explanation for the neutrino mass hierarchy compared to standard seesaw models, which typically require couplings as small as $10^{-6}$ or less. Notably, our approach maintains a relatively simple particle content and does not necessitate additional (gauge) bosons for symmetry breaking.
\end{abstract} 
\maketitle
\newpage

\section{Introduction}

A wide range of theoretical frameworks beyond the Standard Model (BSM) have been proposed to explain neutrino masses.
Among them, numerous ideas have been advanced to address the so-called neutrino hierarchy problem—namely, the challenge of explaining, from the perspective of naturalness, why neutrino masses are extremely small compared to those of other fermion sectors.
One representative example is the radiative seesaw mechanism, typified by the Ma model~\cite{Ma:2006km,Tao:1996vb}. The central philosophy of such approaches is that neutrinos do not directly couple to the SM Higgs boson, but instead interact with new particles such as dark matter candidates. As a result, neutrino masses are generated exclusively at the loop level, acquiring a suppression factor of $\sim1/(4\pi)^2$ per loop, which naturally explains their tiny scale.\\
In addition, models in which neutrino masses originate from suppressed vacuum expectation values (VEVs) have attracted attention as compelling approaches to naturally account for the tiny neutrino mass scale.
Within this context, the simplest construction is the Type-II seesaw model~\cite{Foot:1988aq}, which introduces an SU(2)$_L$ triplet scalar with hypercharge $Y=1$.  Neutrino masses are generated directly through the triplet VEV, $v_\Delta$. 
Precision measurements constrain this VEV to be of order a few GeV, approximately two orders of magnitude smaller than the Standard Model–like Higgs VEV. However, even with such a suppressed $v_\Delta$, explaining the observed neutrino masses requires Yukawa couplings of order $10^{-10}$, which is unsatisfactory from the standpoint of naturalness. 
A natural extension to address this is
to further reduce the VEV in order to enhance the Yukawa couplings, but it is natural to seek a theoretical mechanism that provides a structural explanation rather than relying on ad hoc parameter tuning.

To address this issue, several attempts have been made to induce $v_\Delta$
radiatively by introducing additional symmetries or fields~\cite{Kanemura:2012rj, Okada:2015nca}. Typically, such models impose discrete or continuous symmetries to forbid tree-level contributions. More recently, however, a novel theoretical framework has emerged in which non-invertible symmetries are employed to generate loop effects via dynamical symmetry breaking~\cite{Suzuki:2025oov, Kobayashi:2025cwx, Nomura:2025sod,  Okada:2025kfm, Jangid:2025krp,
Jangid:2025thp, Nomura:2025tvz, Okada:2025adm, Okada:2026gxl, Chen:2025awz}.
\footnote{The other phenomenological applications are discussed in the following references: quark–lepton texture analyses~\cite{Qu:2026omn, Kobayashi:2025ldi, Kobayashi:2024cvp, Kobayashi:2025znw,
Nomura:2025yoa}, strong CP physics~\cite{Liang:2025dkm,Kobayashi:2025thd, Kobayashi:2025rpx}, dark matter models~\cite{Suzuki:2025oov}, suppression of flavor changing neutral currents~\cite{Nakai:2025thw}, realization of family-independent matter symmetries~\cite{Kobayashi:2025lar}, and generalized CP constructions~\cite{Kobayashi:2025wty}.
}
 The advantage of this approach lies in its ability to explain the neutrino mass hierarchy without requiring new (gauge) bosons for symmetry breaking, thereby maintaining a relatively simple particle content.
Despite its promise, the application of non-invertible symmetries raises theoretical challenges, particularly the appearance of divergences in loop calculations. 
To regulate these divergences, a cut-off scale $\Lambda$
has to be introduced. Previous studies~\cite{Nomura:2026hli} have typically adopted phenomenologically reasonable values such as 
$\Lambda\sim10^{5}$ GeV that corresponds to the maximal energy scale accessible at current or near-future experiments such as CEPC-SPPC~\cite{CEPCStudyGroup:2023quu}.
But the absence of a fundamental principle to fix $\Lambda$
has remained a significant obstacle. In cases where the divergence is logarithmic, the weak dependence of results on 
$\Lambda$ has further discouraged detailed discussion. 

 In this work, we propose a scenario in which the cut-off scale 
$\Lambda$
 is determined dynamically from the renormalization group (RGE) flow of the SU(2)$_L$
 gauge coupling $g_2$.
 While in the SM $g_2$ decreases at high energies due to near-asymptotic freedom, the introduction of higher SU(2)$_L$
 multiplets alters the RGE behavior such that $g_2$
grows at high energies and eventually encounters a Landau pole. We define the scale at which $g_2$ reaches the strong coupling regime as the physical cut-off $\Lambda$.
In this way, the dominant contributions to loop integrals arise from the perturbative regime where $g_2$
 remains small, while the upper bound of the integration is naturally fixed by the internal dynamics of the theory, 
 {yielding our framework as an effective field theory valid below $\Lambda$.
 }

In the specific model proposed in this paper, we introduce the following new SU(2)$_L$ multiplet particles:~\footnote{The other intriguing neutrino mass ideas are found in refs.~\cite{Wang:2016lve, Nomura:2018cfu, Nomura:2018lsx, Nomura:2018ktz, Nomura:2018pbw}, introducing such multiplet particles.  }
\begin{itemize}
\item  three families of SU(2)$_L$ quintet fermions $\Sigma_R$ with hypercharge $Y=0$,
\item  an SU(2)$_L$ quartet scalar $\phi_4$ with hypercharge $Y=1/2$.
\end{itemize}
Neutrino masses are generated via a seesaw mechanism mediated by the small VEV $v_4$ of $\phi_4$.
%
%
This VEV is induced at the one-loop level through dynamical symmetry breaking under a non-invertible symmetry governed by the Fibonacci fusion rule. In this construction, the higher multiplets ($\Sigma_R$ and $\phi_4$) play a dual role: they both facilitate the generation of a suppressed $v_4$, thereby explaining the neutrino mass hierarchy, and simultaneously determine the cut-off scale $\Lambda$
 through the RGE flow of $g_2$. As a result, both the experimentally allowed hierarchy $v_4\ll v_H$
and a theoretically motivated finite cut-off scale $\Lambda$
 emerge naturally, without the need to introduce additional (gauge) bosons. Here, $v_H\approx$246 GeV is VEV of the SM Higgs.
{Moreover, these multiplet states are expected to be within the discovery reach of the Large Hadron Collider when their masses lie at the TeV scale, particularly around 1 TeV. A related discussion has previously appeared in Ref.~\cite{Nomura:2017abu}.}

This paper is organized as follows.
In Sec.~II, we introduce our model, derive the active neutrino mass matrix, and discuss the typical order of magnitude of the Yukawa couplings and relevant mass scales.
We conclude and discuss in Sec.~III.

\section{ Model setup}
 \begin{widetext}
\begin{center} 
\begin{table}
\begin{tabular}{|c||c|c|c||c|c|c|}\hline\hline  
&\multicolumn{3}{c||}{Lepton Fields} & \multicolumn{2}{c|}{Scalar Fields} \\\hline
& ~$L_L$~ & ~$e_R$~ & ~$\Sigma_R$ ~ & ~$H$~ & ~$\phi_4$~  \\\hline 
$SU(2)_L$ & $\bm{2}$  & $\bm{1}$  & $\bm{5}$ & $\bm{2}$  & $\bm{4}$  \\\hline 
$U(1)_Y$ & $-\frac12$ & $-1$  & $0$  & $\frac12$ & $\frac12$   \\\hline
${\rm FFR}$ & $\tau$ & $\tau$  & $\tau$  & $\mathbbm{I}$ & $\tau$   \\\hline
\end{tabular}
\caption{Contents of fermion and scalar fields
and their charge assignments under $SU(2)_L\times U(1)_Y$ and Fibonacci fusion rule (FFR).}
\label{tab:1}
\end{table}
\end{center}
\end{widetext}

In this section, we review our model. As we already mentioned in the Introduction,
we introduce three isospin quintet right-handed fermions $\Sigma_R$ which have zero hypercharge,
and add an isospin quartet scalar $\phi_4$ with hypercharge $Y=1/2$ in addition to the SM Higgs denoted by $H$.
Furthermore, we impose $\tau$ to all leptons and bosons except $H$ under the Fibonacci fusion rule (FFR).
 The particle contents and their charges are shown in Tab.~\ref{tab:1}.
{Fibonacci fusion algebra consists of two elements $\mathbbm{I}$ and $\tau$ which is known as the minimal fusion rule.
They satisfy the following relations:
\begin{align}
\label{eq:Fibo}
    \mathbb{I}\otimes \mathbb{I} = \mathbb{I},\quad
    \mathbb{I}\otimes \tau = \tau \otimes \mathbb{I} = \tau,\quad    
    \tau \otimes \tau = \mathbb{I} \oplus \tau.
\end{align}
Following these multiplication rules, $H^\dag\phi_4^* H^T H$ quartic coupling is not allowed by the FFR even but though it is allowed by the $SU(2)_L\otimes U(1)_Y$. In details, we will discuss in the subsection of Higgs potential.
}
We define that neutral components of $H$ and $\phi_4$ have nonzero VEVs, which are respectively denoted by $v_H/\sqrt2$ and $v_4/\sqrt{2}$.
Then,  VEVs are constrained by the following $\rho$ parameter at tree level~\cite{Nomura:2017abu, Nomura:2018cle}:
\begin{align}
\rho=\frac{v_H^2+7 v_4^2}{v_H^2+  v_4^2},
\end{align}
where the experimental bound should be $\rho=1.00038\pm0.00020$ at 1$\sigma$ confidence level~\cite{ParticleDataGroup:2024cfk}.
Combining the deviation of $\rho$ with $v_{\rm SM}=\sqrt{v_H^2+ v_4^2} \simeq 246$ GeV, we find the upper bound as follows:
\begin{align}
v_4 \lesssim  2.8 \ {\rm GeV} , 
\end{align}
where we allow the deviation up to 2$\sigma$ and $v_H\approx 245.984$ GeV.
The relevant Lagrangian under these symmetries is given by
\begin{align}
-\mathcal{L}_{Y}
&=
(y_{\ell})_{ii} \overline{L_{L_i}} H e_{R_i} +(y_{\nu})_{ij} [\overline{L_{L_i}} (i\tau_2) \phi_4^* \Sigma_{R_j} ]
 +  (M_{R})_i [\overline{\Sigma^C_{R_i}} \Sigma_{R_i}] + {\rm h.c.},\label{eq:lep}
\end{align}
where $i=1-3$, $j=1-3$, $\tau_2$ is the second Pauli matrix,  $y_\ell$ and $M_R$ can be taken to be diagonal without loss of generality.
$y_\ell$ generates the SM
charged-lepton mass eigenvalues $m_\ell\equiv y_\ell v_H/\sqrt2$ after the electroweak symmetry breaking.
Note that inside bracket "$[ \ ]$" the $SU(2)_L$ indices are contracted so that it makes singlet where we omit details here.
We work on the basis that all the coefficients are real and positive for simplicity.

\subsection{Fermion quintet}
The quintet Majorana fermions is defined as follows:
\begin{align} 
\Sigma_R&
\equiv
\left[\Sigma_1^{++},\Sigma_1^{+},{\Sigma^0}, \Sigma_{2}^-, \Sigma_{2}^{--} \right]_R^T, 
\label{eq:sigmaR}
\end{align}
where the upper indices for components represent the electric charges while the lower indices distinguish components with the same electric charge. Then, their masses are simply given by
\begin{align}
M_R [\bar \Sigma_R^c \Sigma_R] 
 = M_R
 [\bar \Sigma_{1R}^{++ c} \Sigma_{2R}^{--} + \bar \Sigma_{1R}^{+ c} \Sigma_{2R}^{-} + \bar \Sigma_{2R}^{- c} \Sigma_{1R}^{+} + \bar \Sigma_{2R}^{-- c} \Sigma_{1R}^{++} + 
\bar \Sigma_{R}^{0 c} \Sigma_{R}^{0}].
\end{align}
Thus $\Sigma_{1}^{\pm (\pm \pm)}$ and $\Sigma_{2}^{\pm (\pm \pm)}$ are combined to make singly(doubly)- charged Dirac fermions while $\Sigma^0_R$ remains as the Majorana fermions.
Each of mass is almost degenerate.

\begin{figure}
\centering
\begin{tikzpicture}
  \node[ellipse, draw, dashed,
        minimum width=2.5cm, minimum height=2.5cm] (cluster) {};
  \node (lambda1) at (cluster.west) [xshift=13pt] {$\lambda_{H\phi_4}^{(j)}$};
  \node (lambda2) at (cluster.east) [xshift=-8pt] {$\lambda_{0}^{(i)}$};  
  \node (phi1) at ($(lambda1)+(-1.8cm,1.2cm)$) {$H^\dag$};
  \node (phi2) at ($(lambda1)+(-1.8cm,-1.2cm)$) {$H$};
  \draw[dashed] ($(lambda1)-(8pt,0pt)$) -- (phi1);
  \draw[dashed] ($(lambda1)-(8pt,0pt)$) -- (phi2);
  \node (phi3) at (cluster.north) [yshift=8pt] {$\phi^\dag_4$};
  \node (phi4) at (cluster.south) [yshift=-8pt] {$\phi_4$};
  \draw[dashed] (phi3) -- (cluster.north);
  \draw[dashed] (phi4) -- (cluster.south);
  \node (phi5) at ($(lambda2)+(1.8cm,1.2cm)$) {$\phi_4^*$};
  \node (phi6) at ($(lambda2)+(1.8cm,-1.2cm)$) {$H^T$};
  \draw[dashed] ($(lambda2)+(8pt,0cm)$) -- (phi5);
  \draw[dashed] ($(lambda2)+(8pt,0cm)$) -- (phi6);
\end{tikzpicture}
\caption{One-loop diagram of  the quartic term $\delta\lambda_{ij}$.}
\label{fig:dlam}
\end{figure}
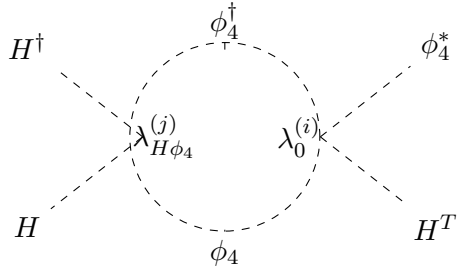

\subsection{Higgs potential}
The scalar fields can be parameterized as 
\begin{align}
&H =\left[
\begin{array}{c}
w^+\\
\frac{v_H+h+iz}{\sqrt2}
\end{array}\right],\quad 
\phi_4 =\left[\varphi^{++}, \varphi_2^{+}, \varphi^0,\varphi_1^{-}\right]^T,
\label{component}
\end{align}
where the upper indices of each component represents the electric charges while the lower indices distinguish components with same electric charge, and $\varphi^0\equiv \frac{v_4+\varphi_R+i\varphi_I}{\sqrt2}$.
%
Then, the renormalizable potential $\mathcal{V}= \mathcal{V}_2+\mathcal{V}_4$ under our symmetries is given by
\begin{align}
\mathcal{V}_2 &= -\mu_H^2 H^\dagger H + \mu_4^2 \phi_4^\dagger \phi_4,\\
 \mathcal{V}_4&= \lambda_0^{(i)} [\phi_4^\dag H  \phi_4^\dag \phi_4]_i 
+ \lambda_{H\phi_4}^{(i)} [H^\dag H \phi_4^\dag   \phi_4]_i +{\rm c.c.} \label{eq:pot4-1} \\
&
+ \lambda_{H} |H|^4 + \lambda_{\phi_4}^{(i)} [\phi_4^\dag   \phi_4\phi_4^\dag   \phi_4]_i
\label{Eq:lag-flavor}
\end{align}
where $i=1,2$, $\tilde\Phi\equiv i\tau_2\Phi^*$, we consider $\mathcal{V}_2$ almost corresponds to mass eigenvalues for each component.
{\it  
Here, we stress to mention that $\delta\lambda_{ij} [H^\dag\phi_4^* H^T H]_{ij}$ term is generated at one-loop level through $\lambda_0^{(i)} [\phi_4^\dag H  \phi_4^\dag \phi_4]_i $ and $ \lambda_{H\phi_4}^{(i)} [H^\dag H \phi_4^\dag   \phi_4]_i$ in Eq.~\ref{eq:pot4-1}
{as shown in Fig.~\ref{fig:dlam}.}
Clearly, this term dynamically violates FFR and $\delta\lambda$ is given by 
\begin{align}
\delta\lambda_{ij} \approx 
- \frac{\lambda_0^{(i)} \lambda_{H\phi_4}^{(j)}}{(4\pi)^2}
\left[
\frac{2-2r+(1+r)\ln(r)}{1-r}
\right],
\end{align}
where $r\equiv m_H^2/\Lambda^2$, $\Lambda$ being our cut-off scale which we discussed in Introduction.
}
Then, $v_H\approx \mu_H/\sqrt{\lambda_H}$, and $v_4$ is approximately given by~\cite{Kanemura:2013qva}
\begin{align}
v_4&\sim  v_H^3 \sum_{i,j=1}^2  \frac{\delta\lambda_{ij}  }{ 2\mu_4^2 +(\lambda_{H\phi_4}^{(1)} +\lambda_{H\phi_4}^{(2)}) v_H^2}\\
&=
-\frac{v_H^3}{(4\pi)^2}
\left[\frac{2-2r+(1+r)\ln(r)}{1-r} \right]
 \sum_{i,j=1}^2  \frac{ \lambda_0^{(i)} \lambda_{H\phi_4}^{(j)}}{ 2\mu_4^2 +(\lambda_{H\phi_4}^{(1)} +\lambda_{H\phi_4}^{(2)}) v_H^2}
\\
&\simeq
-\frac{v_H}{2(4\pi)^2} \frac{v^2_H}{\mu_4^2}
\left[\frac{2-2r+(1+r)\ln(r)}{1-r} \right] 
 \sum_{i,j=1}^2   \lambda_0^{(i)} \lambda_{H\phi_4}^{(j)} 
 ,
\end{align}
where $\lambda_{H\phi_4}^{(1)} +\lambda_{H\phi_4}^{(2)}\ll \mu_4^2/v_H^2$ in the last line.
Now we need to know the concrete energy scale of $\Lambda$ ($r\equiv m^2_H/\Lambda^2$) in order to estimate the value of $v_4$,
we have to evaluate the renormalization group evolution specifically on $g_2$ before discussing the neutrino mass matrix.
{Here, we briefly comment on the dependence of $v_4$ on the cut-off scale.
For instance, when the cut-off is rescaled as $\Lambda\to 2\Lambda$ with e.g. $\Lambda=100$ TeV, the ratio $v_4(2\Lambda)/v_4(\Lambda)$ is approximately 1.22.
Therefore, the sensitivity of $v_4$ to the cut-off scale is negligibly small.
}

\begin{figure}[tb]
\begin{center}
\includegraphics[width=13cm]{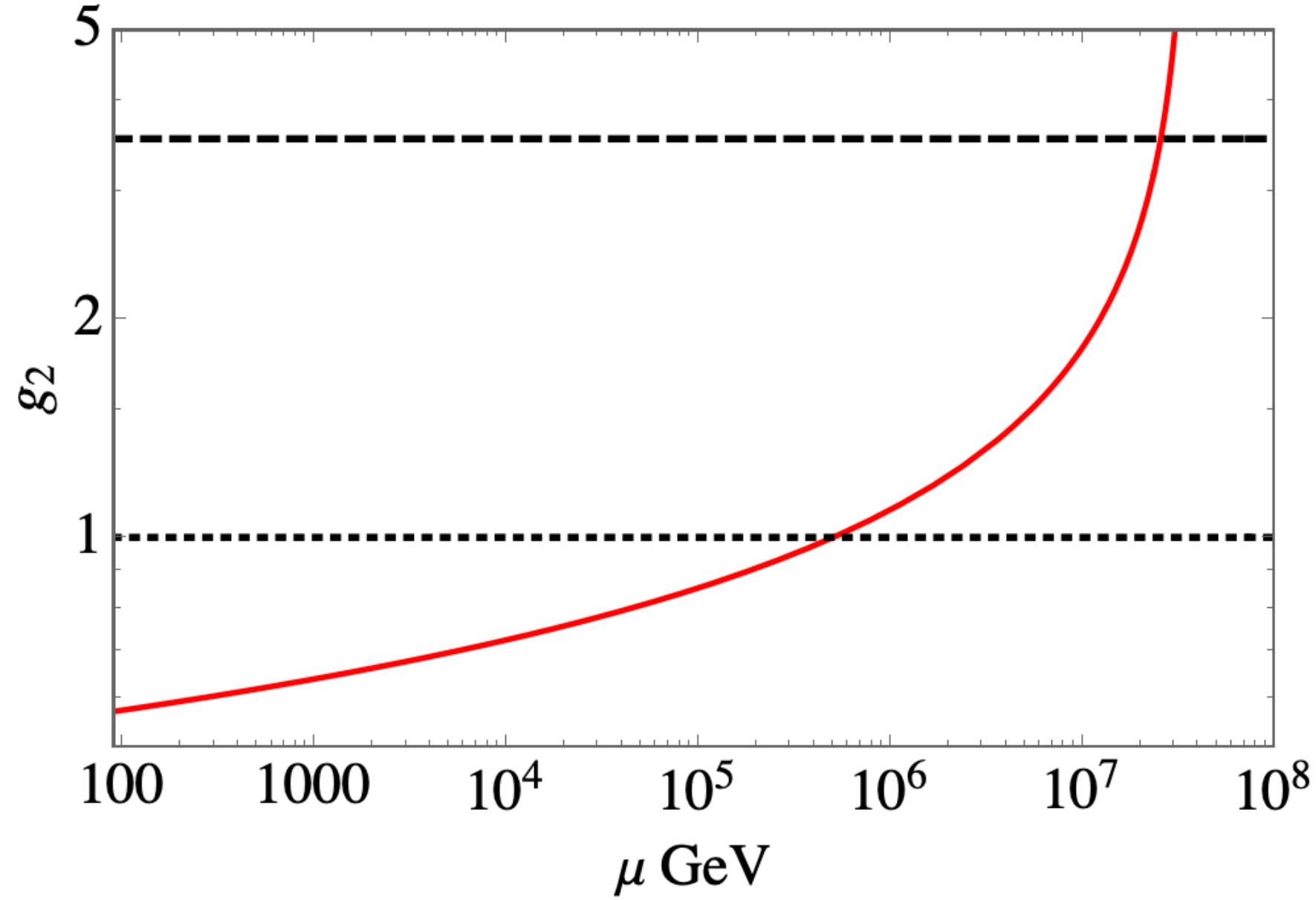}
\caption{The running of $g_2$ in terms of a reference energy of $\mu$, where dashed (dotted) horizontal line represents $g_2=\sqrt{4\pi}\ (1)$.}
\label{fig:rge}
\end{center}\end{figure}

\subsection{Beta function of $g_2$}
\label{beta-func}
Here we discuss running of the SU(2)$_L$ gauge coupling $g_2$ in the presence of new particles $\Sigma_R$ and $\phi_4$.
~\footnote{Note here that U(1)$_Y$ gauge coupling $g_Y$ runs far beyond the Planck scale without divergence.}
The new beta function of $g_2$ for three $SU(2)_L$ quintet fermions ($\Sigma_R$) and a quartet boson ($\Phi_4$) is given by
\begin{align}
\Delta b^{\Sigma_R}_g=3\times \frac{20}{3}, \quad \Delta b^{\Phi_4}_g=\frac{5}{3} .
\end{align}
Then, the energy evolution of the gauge coupling $g_2$ as~\cite{Kanemura:2015bli}
\begin{align}
\frac{1}{g_2^2(\mu)}&=\frac1{g_2^2(m_{in.})}-\frac{b^{SM}_g}{(4\pi)^2}\ln\left[\frac{\mu^2}{m_{in.}^2}\right]\nn\\
&
-\theta(\mu-m_{th.}) \frac{\Delta b^{\Sigma_R}_g}{(4\pi)^2}\ln\left[\frac{\mu^2}{m_{th.}^2}\right]
-\theta(\mu-m_{th.}) \frac{\Delta b^{\Phi_4}_g}{(4\pi)^2}\ln\left[\frac{\mu^2}{m_{th.}^2}\right],\label{eq:rge_g}
\end{align}
where $\mu$ is a reference energy, $b^{SM}_g=-19/6$, and we suppose to be $m_{in.}(=m_Z)<m_{th.}=$1000 GeV, being respectively threshold masses of $\Sigma_R$ and $\phi_4$ for $m_{th.}$.
The resultant energy flow of $g_2(\mu)$ is given by the Fig.~\ref{fig:rge}, where dashed (dotted) horizontal line represents $g_2=\sqrt{4\pi}\ (1)$.
This figure shows that $g_2$ is relevant up to the mass scale $\mu\simeq2.56\times10^{7}(5.05\times10^5)$ GeV when $g_2$ should be within $\sqrt{4\pi}(1)$.
Thus, we identify this upper mass scale of $\mu$ to be our cut-off scale $\Lambda$.
\\
\underline{{\bf NOTE on RGE:}}
{
Before proceeding with our discussion, it is important to address certain issues related to the renormalization group equations (RGEs)~\cite{Machacek:1983tz, Machacek:1983fi, Machacek:1984zw}. When a large number of  SU(2)$_L$ multiplets are introduced in a model, one might naively expect that the cutoff scale is determined solely by the divergence of the gauge coupling $g_2$, since it typically runs most rapidly toward strong coupling. However, this is not always the case. 
In particular, the self-quartic coupling of the multiplet boson ($\phi_4$ in our case) denoted $\lambda_{\phi_4}^{(i)}\ (i=1,2)$, may diverge at an energy scale lower than or comparable to the divergence scale of $g_2$.\\
%
%
%
%
%
%
To ensure that the physical cut-off $\Lambda$ is uniquely and consistently determined by the $SU(2)_L$ gauge sector, the initial values of the scalar quartic couplings at the electroweak scale must be sufficiently suppressed, e.g., $\lambda^{(i)}_{\phi_4}(\mu)/g^2_2(\mu)\ll 1$  $(i=1,2)$.
This choice prevents the scalar sector from reaching a Landau pole prematurely, allowing the theory to remain under perturbative control until it encounters the gauge-induced singularity.
Moreover, the neutrino Yukawa coupling $y_\nu$ can also influence the RGE flow of $\lambda_4$ if $y_\nu$ is taken too large.
We follow the requirement $y_\nu \lesssim 0.2$ to minimize the Yukawa contribution to the RGEs.
In this case, the contribution of $y_\nu$ to the running of $\lambda_4$ remains below 10\% of the gauge contribution from $g_2$,
and the system remains under control.
A full analysis of the  RGE behaviors for all couplings lies beyond the scope of our work, but the above estimates suffice to determine the cut-off scale relevant to our discussion.

\subsection{ Neutrino mass matrix}
The neutrino mass matrix is given through $(y_{\nu})_{ij} [\overline{L_{L_i}} (i\tau_2) \phi_4^* \Sigma_{R_j} ]$ and $M_R[\bar \Sigma_{R}^C \Sigma_{R}]$ with the seesaw mechanism. The relevant component is found as
\begin{align}
-{\cal L}& \supset  \frac{(y_\nu)_{ia}}{\sqrt2}
\bar\nu_{L_i} \Sigma_{R_a}^0 \varphi^{0*} + M_{R_a} \overline{\Sigma_{R_a}^C} \Sigma_{R_a}
 +{\rm h.c.}.
\label{eq:Yukawa}
\end{align}

After spontaneous symmetry breaking of $\phi_4$, 
the active neutrino mass matrix $m_\nu$
is given by 
\begin{align}
& (m_{\nu})_{ij}
=v_4^2 \sum_{a=1}^3{(y_\nu)_{ia}M_{R_a}^{-1} (y_\nu^T)_{aj}} .
\end{align}
Then, the  mass matrix $m_\nu$ is diagonalized by an observed unitary mixing matrix $V$~\cite{Maki:1962mu} as
$D_\nu(\equiv {\rm diag}[D_1,D_2,D_3])=V^\dag m_\nu V^*$, where we adopt $V$ for the standard parameterization. 
\begin{align}
V=
\left[\begin{array}{ccc} {c_{13}}c_{12} &c_{13}s_{12} & s_{13} e^{-i\delta}\\
 -c_{23}s_{12}-s_{23}s_{13}c_{12}e^{i\delta} & c_{23}c_{12}-s_{23}s_{13}s_{12}e^{i\delta} & s_{23}c_{13}\\
  s_{23}s_{12}-c_{23}s_{13}c_{12}e^{i\delta} & -s_{23}c_{12}-c_{23}s_{13}s_{12}e^{i\delta} & c_{23}c_{13}\\
  \end{array}
\right]
\left[\begin{array}{ccc} 1 &0 & 0 \\
0 & e^{i\alpha/2} & 0 \\
0 &0 & e^{i\beta/2} \\  \end{array}
\right],
\end{align}
where $\delta$ is Dirac CP phase, and $\alpha,\ \beta$ are Majorana phases.
Two mass squared differences, three mixing angles, and $\delta$ are experimentally obtained in Nufit 6.1~\cite{Esteban:2024eli}. 
Then, we can rewrite the Yukawa coupling $y_\nu$ in terms of observed values and some of the free parameters, so-called Casas-Ibarra parametrization~\cite{Casas:2001sr}, as follows:
\begin{align}
y_\nu
&= \frac1{v_4} V \sqrt{D_\nu} O \sqrt{M_R},
\label{eq:ci}
\end{align}
where $O$ is an arbitrary complex orthogonal matrix with three degrees of freedom; $OO^T=O^TO=1_{3\times3}$.
Here we estimate $v_4$.
When $\Lambda\sim 2.56\times10^{7} (5.05\times10^5)$ GeV,
$v_4$ is given by
\begin{align}
&\Lambda\sim 2.56\times10^{7}\ {\rm GeV}:\quad 
v_4 \sim 11.23 \frac{v_H}{(4\pi)^2} \frac{v^2_H}{\mu_4^2}  \sum_{i,j=1}^2   \lambda_0^{(i)} \lambda_{H\phi_4}^{(j)} ,\\
&\Lambda\sim 5.05\times10^5\ {\rm GeV}:\quad 
v_4 \sim 7.30 \frac{v_H}{(4\pi)^2} \frac{v^2_H}{\mu_4^2}  \sum_{i,j=1}^2   \lambda_0^{(i)} \lambda_{H\phi_4}^{(j)} .
\end{align}
Furthermore, if $ \sum_{i,j=1}^2   \lambda_0^{(i)} \lambda_{H\phi_4}^{(j)} \sim 0.1$ and $\mu_4\sim1000$ GeV, 
one obtains the following $v_4$: 
 \begin{align}
&\Lambda\sim 2.56\times10^{7}\ {\rm GeV}:\quad 
v_4 \sim 0.1058\ {\rm GeV},\\
&\Lambda\sim 5.05\times10^5\ {\rm GeV}:\quad 
v_4 \sim 0.0688\ {\rm GeV} .
\end{align}
Then, $y_\nu$  can be estimated under the 
assumption $|V|\approx |O| =1$ and $D_\nu\sim 100$ meV, $M_R\sim 1000$ GeV, and we obtain the following coupling
  \begin{align}
&\Lambda\sim 2.56\times10^{7}\ {\rm GeV}:\quad 
y_\nu \sim 3\times10^{-3},\\
&\Lambda\sim 5.05\times10^5\ {\rm GeV}:\quad 
y_\nu \sim 4.6\times10^{-3}.
\end{align}
The typical seesaw model requires $y_\nu\sim 10^{-6}$ under the same condition,  our scenario is relaxed by three orders of magnitude!

\section{ Conclusions and discussions}
 In this paper, we have proposed 
 {our framework as an effective field theory valid below the cut-off scale $\Lambda$}
 to explain the tiny scale of neutrino masses by integrating non-invertible symmetries with the dynamical determination of the cut-off scale. Our model introduces three families of SU(2)$_L$
quintet fermions ($\Sigma_R$) and a quartet scalar $(\phi_4)$,
where neutrino masses are generated via a seesaw mechanism mediated by the suppressed vacuum expectation value $v_4$
of the quartet.
The core achievement of this work is the resolution of the arbitrariness of the cut-off scale $\Lambda$ in loop-induced VEV models.
We have demonstrated that the introduction of higher $SU(2)_L$ multiplets alters the renormalization evolution of the $SU(2)_L$
gauge coupling $g_2$, causing it to encounter a Landau pole at high energies.
By identifying the scale where $g_2$
reaches the strong coupling regime as the physical cut-off $\Lambda$, we eliminate the need to introduce this scale by hand. Our analysis shows that $\Lambda$ is naturally determined to be in the range of approximately $10^5$ to $10^7$ GeV.
Quantitatively, this framework successfully explains the neutrino mass hierarchy without extreme parameter tuning.
Under the Fibonacci fusion rule, the $v_4$ is induced at the one-loop level.
For a cut-off scale $\Lambda=10^5\sim10^7$ GeV, we obtained $v_4=0.07\sim0.1$ GeV.
Consequently, the required neutrino Yukawa couplings are on the order of $10^{-3}$,
which is significantly larger and thus more natural than the $10^{-6}$ to $10^{-10}$ typically required in the standard seesaw or Type-II models.
In conclusion, our model provides a structural explanation for the smallness of neutrino masses by linking the internal dynamics of gauge
coupling evolution to the generation of loop-induced scales. This approach maintains a relatively simple particle content without requiring additional gauge bosons for symmetry breaking. Future studies could further explore the phenomenological implications at the LHC, particularly the pair production of charged particles within the quintet and quartet multiplets, as well as a more comprehensive RGE analysis including all scalar quartic couplings.

\section*{Acknowledgments}
\vspace{0.5cm}
H. O. is supported by the Zhongyuan Talent (Talent Recruitment Series) Foreign Experts Project.
\bibliography{references.bib}

\begin{thebibliography}{44}
\expandafter\ifx\csname natexlab\endcsname\relax\def\natexlab#1{#1}\fi
\expandafter\ifx\csname bibnamefont\endcsname\relax
  \def\bibnamefont#1{#1}\fi
\expandafter\ifx\csname bibfnamefont\endcsname\relax
  \def\bibfnamefont#1{#1}\fi
\expandafter\ifx\csname citenamefont\endcsname\relax
  \def\citenamefont#1{#1}\fi
\expandafter\ifx\csname url\endcsname\relax
  \def\url#1{\texttt{#1}}\fi
\expandafter\ifx\csname urlprefix\endcsname\relax\def\urlprefix{URL }\fi
\providecommand{\bibinfo}[2]{#2}
\providecommand{\eprint}[2][]{\url{#2}}

\bibitem[{\citenamefont{Ma}(2006)}]{Ma:2006km}
\bibinfo{author}{\bibfnamefont{E.}~\bibnamefont{Ma}}, \bibinfo{journal}{Phys.
  Rev. D} \textbf{\bibinfo{volume}{73}}, \bibinfo{pages}{077301}
  (\bibinfo{year}{2006}), \eprint{hep-ph/0601225}.

\bibitem[{\citenamefont{Tao}(1996)}]{Tao:1996vb}
\bibinfo{author}{\bibfnamefont{Z.-j.} \bibnamefont{Tao}},
  \bibinfo{journal}{Phys. Rev. D} \textbf{\bibinfo{volume}{54}},
  \bibinfo{pages}{5693} (\bibinfo{year}{1996}), \eprint{hep-ph/9603309}.

\bibitem[{\citenamefont{Foot et~al.}(1989)\citenamefont{Foot, Lew, He, and
  Joshi}}]{Foot:1988aq}
\bibinfo{author}{\bibfnamefont{R.}~\bibnamefont{Foot}},
  \bibinfo{author}{\bibfnamefont{H.}~\bibnamefont{Lew}},
  \bibinfo{author}{\bibfnamefont{X.~G.} \bibnamefont{He}}, \bibnamefont{and}
  \bibinfo{author}{\bibfnamefont{G.~C.} \bibnamefont{Joshi}},
  \bibinfo{journal}{Z. Phys. C} \textbf{\bibinfo{volume}{44}},
  \bibinfo{pages}{441} (\bibinfo{year}{1989}).

\bibitem[{\citenamefont{Kanemura and Sugiyama}(2012)}]{Kanemura:2012rj}
\bibinfo{author}{\bibfnamefont{S.}~\bibnamefont{Kanemura}} \bibnamefont{and}
  \bibinfo{author}{\bibfnamefont{H.}~\bibnamefont{Sugiyama}},
  \bibinfo{journal}{Phys. Rev. D} \textbf{\bibinfo{volume}{86}},
  \bibinfo{pages}{073006} (\bibinfo{year}{2012}), \eprint{1202.5231}.

\bibitem[{\citenamefont{Okada and Orikasa}(2016)}]{Okada:2015nca}
\bibinfo{author}{\bibfnamefont{H.}~\bibnamefont{Okada}} \bibnamefont{and}
  \bibinfo{author}{\bibfnamefont{Y.}~\bibnamefont{Orikasa}},
  \bibinfo{journal}{Phys. Rev. D} \textbf{\bibinfo{volume}{93}},
  \bibinfo{pages}{013008} (\bibinfo{year}{2016}), \eprint{1509.04068}.

\bibitem[{\citenamefont{Suzuki and Xu}(2025)}]{Suzuki:2025oov}
\bibinfo{author}{\bibfnamefont{M.}~\bibnamefont{Suzuki}} \bibnamefont{and}
  \bibinfo{author}{\bibfnamefont{L.-X.} \bibnamefont{Xu}}
  (\bibinfo{year}{2025}), \eprint{2503.19964}.

\bibitem[{\citenamefont{Kobayashi
  et~al.}(2025{\natexlab{a}})\citenamefont{Kobayashi, Okada, and
  Otsuka}}]{Kobayashi:2025cwx}
\bibinfo{author}{\bibfnamefont{T.}~\bibnamefont{Kobayashi}},
  \bibinfo{author}{\bibfnamefont{H.}~\bibnamefont{Okada}}, \bibnamefont{and}
  \bibinfo{author}{\bibfnamefont{H.}~\bibnamefont{Otsuka}}
  (\bibinfo{year}{2025}{\natexlab{a}}), \eprint{2505.14878}.

\bibitem[{\citenamefont{Nomura and Okada}(2025)}]{Nomura:2025sod}
\bibinfo{author}{\bibfnamefont{T.}~\bibnamefont{Nomura}} \bibnamefont{and}
  \bibinfo{author}{\bibfnamefont{H.}~\bibnamefont{Okada}}
  (\bibinfo{year}{2025}), \eprint{2506.16706}.

\bibitem[{\citenamefont{Okada and Shigekami}(2025)}]{Okada:2025kfm}
\bibinfo{author}{\bibfnamefont{H.}~\bibnamefont{Okada}} \bibnamefont{and}
  \bibinfo{author}{\bibfnamefont{Y.}~\bibnamefont{Shigekami}}
  (\bibinfo{year}{2025}), \eprint{2507.16198}.

\bibitem[{\citenamefont{Jangid and Okada}(2025{\natexlab{a}})}]{Jangid:2025krp}
\bibinfo{author}{\bibfnamefont{S.}~\bibnamefont{Jangid}} \bibnamefont{and}
  \bibinfo{author}{\bibfnamefont{H.}~\bibnamefont{Okada}}
  (\bibinfo{year}{2025}{\natexlab{a}}), \eprint{2508.16174}.

\bibitem[{\citenamefont{Jangid and Okada}(2025{\natexlab{b}})}]{Jangid:2025thp}
\bibinfo{author}{\bibfnamefont{S.}~\bibnamefont{Jangid}} \bibnamefont{and}
  \bibinfo{author}{\bibfnamefont{H.}~\bibnamefont{Okada}}
  (\bibinfo{year}{2025}{\natexlab{b}}), \eprint{2510.17292}.

\bibitem[{\citenamefont{Nomura et~al.}(2025)\citenamefont{Nomura, Okada, and
  Shigekami}}]{Nomura:2025tvz}
\bibinfo{author}{\bibfnamefont{T.}~\bibnamefont{Nomura}},
  \bibinfo{author}{\bibfnamefont{H.}~\bibnamefont{Okada}}, \bibnamefont{and}
  \bibinfo{author}{\bibfnamefont{Y.}~\bibnamefont{Shigekami}}
  (\bibinfo{year}{2025}), \eprint{2510.17156}.

\bibitem[{\citenamefont{Okada and Shoji}(2025)}]{Okada:2025adm}
\bibinfo{author}{\bibfnamefont{H.}~\bibnamefont{Okada}} \bibnamefont{and}
  \bibinfo{author}{\bibfnamefont{Y.}~\bibnamefont{Shoji}}
  (\bibinfo{year}{2025}), \eprint{2512.20891}.

\bibitem[{\citenamefont{Okada and Shigekami}(2026)}]{Okada:2026gxl}
\bibinfo{author}{\bibfnamefont{H.}~\bibnamefont{Okada}} \bibnamefont{and}
  \bibinfo{author}{\bibfnamefont{Y.}~\bibnamefont{Shigekami}}
  (\bibinfo{year}{2026}), \eprint{2601.15749}.

\bibitem[{\citenamefont{Chen et~al.}(2025)\citenamefont{Chen, Geng, Okada, and
  Wu}}]{Chen:2025awz}
\bibinfo{author}{\bibfnamefont{J.}~\bibnamefont{Chen}},
  \bibinfo{author}{\bibfnamefont{C.-Q.} \bibnamefont{Geng}},
  \bibinfo{author}{\bibfnamefont{H.}~\bibnamefont{Okada}}, \bibnamefont{and}
  \bibinfo{author}{\bibfnamefont{J.-J.} \bibnamefont{Wu}}
  (\bibinfo{year}{2025}), \eprint{2507.11951}.

\bibitem[{\citenamefont{Qu et~al.}(2026)\citenamefont{Qu, Jiang, and
  Ding}}]{Qu:2026omn}
\bibinfo{author}{\bibfnamefont{B.-Y.} \bibnamefont{Qu}},
  \bibinfo{author}{\bibfnamefont{Z.}~\bibnamefont{Jiang}}, \bibnamefont{and}
  \bibinfo{author}{\bibfnamefont{G.-J.} \bibnamefont{Ding}}
  (\bibinfo{year}{2026}), \eprint{2602.24214}.

\bibitem[{\citenamefont{Kobayashi
  et~al.}(2025{\natexlab{b}})\citenamefont{Kobayashi, Otsuka, Tanimoto, and
  Uchida}}]{Kobayashi:2025ldi}
\bibinfo{author}{\bibfnamefont{T.}~\bibnamefont{Kobayashi}},
  \bibinfo{author}{\bibfnamefont{H.}~\bibnamefont{Otsuka}},
  \bibinfo{author}{\bibfnamefont{M.}~\bibnamefont{Tanimoto}}, \bibnamefont{and}
  \bibinfo{author}{\bibfnamefont{H.}~\bibnamefont{Uchida}}
  (\bibinfo{year}{2025}{\natexlab{b}}), \eprint{2505.07262}.

\bibitem[{\citenamefont{Kobayashi et~al.}(2024)\citenamefont{Kobayashi, Otsuka,
  and Tanimoto}}]{Kobayashi:2024cvp}
\bibinfo{author}{\bibfnamefont{T.}~\bibnamefont{Kobayashi}},
  \bibinfo{author}{\bibfnamefont{H.}~\bibnamefont{Otsuka}}, \bibnamefont{and}
  \bibinfo{author}{\bibfnamefont{M.}~\bibnamefont{Tanimoto}},
  \bibinfo{journal}{JHEP} \textbf{\bibinfo{volume}{12}}, \bibinfo{pages}{117}
  (\bibinfo{year}{2024}), \eprint{2409.05270}.

\bibitem[{\citenamefont{Kobayashi
  et~al.}(2025{\natexlab{c}})\citenamefont{Kobayashi, Nishioka, Otsuka, and
  Tanimoto}}]{Kobayashi:2025znw}
\bibinfo{author}{\bibfnamefont{T.}~\bibnamefont{Kobayashi}},
  \bibinfo{author}{\bibfnamefont{Y.}~\bibnamefont{Nishioka}},
  \bibinfo{author}{\bibfnamefont{H.}~\bibnamefont{Otsuka}}, \bibnamefont{and}
  \bibinfo{author}{\bibfnamefont{M.}~\bibnamefont{Tanimoto}},
  \bibinfo{journal}{JHEP} \textbf{\bibinfo{volume}{05}}, \bibinfo{pages}{177}
  (\bibinfo{year}{2025}{\natexlab{c}}), \eprint{2503.09966}.

\bibitem[{\citenamefont{Nomura and Popov}(2025)}]{Nomura:2025yoa}
\bibinfo{author}{\bibfnamefont{T.}~\bibnamefont{Nomura}} \bibnamefont{and}
  \bibinfo{author}{\bibfnamefont{O.}~\bibnamefont{Popov}}
  (\bibinfo{year}{2025}), \eprint{2507.10299}.

\bibitem[{\citenamefont{Liang and Yanagida}(2025)}]{Liang:2025dkm}
\bibinfo{author}{\bibfnamefont{Q.}~\bibnamefont{Liang}} \bibnamefont{and}
  \bibinfo{author}{\bibfnamefont{T.~T.} \bibnamefont{Yanagida}}
  (\bibinfo{year}{2025}), \eprint{2505.05142}.

\bibitem[{\citenamefont{Kobayashi
  et~al.}(2025{\natexlab{d}})\citenamefont{Kobayashi, Otsuka, and
  Yanagida}}]{Kobayashi:2025thd}
\bibinfo{author}{\bibfnamefont{T.}~\bibnamefont{Kobayashi}},
  \bibinfo{author}{\bibfnamefont{H.}~\bibnamefont{Otsuka}}, \bibnamefont{and}
  \bibinfo{author}{\bibfnamefont{T.~T.} \bibnamefont{Yanagida}}
  (\bibinfo{year}{2025}{\natexlab{d}}), \eprint{2508.12287}.

\bibitem[{\citenamefont{Kobayashi
  et~al.}(2025{\natexlab{e}})\citenamefont{Kobayashi, Otsuka, Tanimoto, and
  Yanagida}}]{Kobayashi:2025rpx}
\bibinfo{author}{\bibfnamefont{T.}~\bibnamefont{Kobayashi}},
  \bibinfo{author}{\bibfnamefont{H.}~\bibnamefont{Otsuka}},
  \bibinfo{author}{\bibfnamefont{M.}~\bibnamefont{Tanimoto}}, \bibnamefont{and}
  \bibinfo{author}{\bibfnamefont{T.~T.} \bibnamefont{Yanagida}}
  (\bibinfo{year}{2025}{\natexlab{e}}), \eprint{2510.01680}.

\bibitem[{\citenamefont{Nakai et~al.}(2025)\citenamefont{Nakai, Otsuka,
  Shigekami, and Zhang}}]{Nakai:2025thw}
\bibinfo{author}{\bibfnamefont{Y.}~\bibnamefont{Nakai}},
  \bibinfo{author}{\bibfnamefont{H.}~\bibnamefont{Otsuka}},
  \bibinfo{author}{\bibfnamefont{Y.}~\bibnamefont{Shigekami}},
  \bibnamefont{and} \bibinfo{author}{\bibfnamefont{Z.}~\bibnamefont{Zhang}}
  (\bibinfo{year}{2025}), \eprint{2512.21509}.

\bibitem[{\citenamefont{Kobayashi
  et~al.}(2025{\natexlab{f}})\citenamefont{Kobayashi, Mita, Otsuka, and
  Sakuma}}]{Kobayashi:2025lar}
\bibinfo{author}{\bibfnamefont{T.}~\bibnamefont{Kobayashi}},
  \bibinfo{author}{\bibfnamefont{H.}~\bibnamefont{Mita}},
  \bibinfo{author}{\bibfnamefont{H.}~\bibnamefont{Otsuka}}, \bibnamefont{and}
  \bibinfo{author}{\bibfnamefont{R.}~\bibnamefont{Sakuma}}
  (\bibinfo{year}{2025}{\natexlab{f}}), \eprint{2506.10241}.

\bibitem[{\citenamefont{Kobayashi and Otsuka}(2025)}]{Kobayashi:2025wty}
\bibinfo{author}{\bibfnamefont{T.}~\bibnamefont{Kobayashi}} \bibnamefont{and}
  \bibinfo{author}{\bibfnamefont{H.}~\bibnamefont{Otsuka}}
  (\bibinfo{year}{2025}), \eprint{2512.16376}.

\bibitem[{\citenamefont{Nomura et~al.}(2026)\citenamefont{Nomura, Okada, and
  Shigekami}}]{Nomura:2026hli}
\bibinfo{author}{\bibfnamefont{T.}~\bibnamefont{Nomura}},
  \bibinfo{author}{\bibfnamefont{H.}~\bibnamefont{Okada}}, \bibnamefont{and}
  \bibinfo{author}{\bibfnamefont{Y.}~\bibnamefont{Shigekami}}
  (\bibinfo{year}{2026}), \eprint{2603.15382}.

\bibitem[{\citenamefont{Abdallah et~al.}(2024)}]{CEPCStudyGroup:2023quu}
\bibinfo{author}{\bibfnamefont{W.}~\bibnamefont{Abdallah}} \bibnamefont{et~al.}
  (\bibinfo{collaboration}{CEPC Study Group}), \bibinfo{journal}{Radiat.
  Detect. Technol. Methods} \textbf{\bibinfo{volume}{8}}, \bibinfo{pages}{1}
  (\bibinfo{year}{2024}), \bibinfo{note}{[Erratum:
  Radiat.Detect.Technol.Methods 9, 184--192 (2025)]}, \eprint{2312.14363}.

\bibitem[{\citenamefont{Wang and Han}(2017)}]{Wang:2016lve}
\bibinfo{author}{\bibfnamefont{W.}~\bibnamefont{Wang}} \bibnamefont{and}
  \bibinfo{author}{\bibfnamefont{Z.-L.} \bibnamefont{Han}},
  \bibinfo{journal}{JHEP} \textbf{\bibinfo{volume}{04}}, \bibinfo{pages}{166}
  (\bibinfo{year}{2017}), \eprint{1611.03240}.

\bibitem[{\citenamefont{Nomura and Okada}(2019{\natexlab{a}})}]{Nomura:2018cfu}
\bibinfo{author}{\bibfnamefont{T.}~\bibnamefont{Nomura}} \bibnamefont{and}
  \bibinfo{author}{\bibfnamefont{H.}~\bibnamefont{Okada}},
  \bibinfo{journal}{Phys. Rev. D} \textbf{\bibinfo{volume}{99}},
  \bibinfo{pages}{055027} (\bibinfo{year}{2019}{\natexlab{a}}),
  \eprint{1807.04555}.

\bibitem[{\citenamefont{Nomura and Okada}(2019{\natexlab{b}})}]{Nomura:2018lsx}
\bibinfo{author}{\bibfnamefont{T.}~\bibnamefont{Nomura}} \bibnamefont{and}
  \bibinfo{author}{\bibfnamefont{H.}~\bibnamefont{Okada}},
  \bibinfo{journal}{Phys. Dark Univ.} \textbf{\bibinfo{volume}{26}},
  \bibinfo{pages}{100359} (\bibinfo{year}{2019}{\natexlab{b}}),
  \eprint{1808.05476}.

\bibitem[{\citenamefont{Nomura and Okada}(2019{\natexlab{c}})}]{Nomura:2018ktz}
\bibinfo{author}{\bibfnamefont{T.}~\bibnamefont{Nomura}} \bibnamefont{and}
  \bibinfo{author}{\bibfnamefont{H.}~\bibnamefont{Okada}},
  \bibinfo{journal}{Phys. Lett. B} \textbf{\bibinfo{volume}{792}},
  \bibinfo{pages}{424} (\bibinfo{year}{2019}{\natexlab{c}}),
  \eprint{1809.06039}.

\bibitem[{\citenamefont{Nomura and Okada}(2019{\natexlab{d}})}]{Nomura:2018pbw}
\bibinfo{author}{\bibfnamefont{T.}~\bibnamefont{Nomura}} \bibnamefont{and}
  \bibinfo{author}{\bibfnamefont{H.}~\bibnamefont{Okada}},
  \bibinfo{journal}{Nucl. Phys.} \textbf{\bibinfo{volume}{B}},
  \bibinfo{pages}{114621} (\bibinfo{year}{2019}{\natexlab{d}}),
  \eprint{1812.08016}.

\bibitem[{\citenamefont{Nomura and Okada}(2017)}]{Nomura:2017abu}
\bibinfo{author}{\bibfnamefont{T.}~\bibnamefont{Nomura}} \bibnamefont{and}
  \bibinfo{author}{\bibfnamefont{H.}~\bibnamefont{Okada}},
  \bibinfo{journal}{Phys. Rev. D} \textbf{\bibinfo{volume}{96}},
  \bibinfo{pages}{095017} (\bibinfo{year}{2017}), \eprint{1708.03204}.

\bibitem[{\citenamefont{Nomura and Okada}(2018)}]{Nomura:2018cle}
\bibinfo{author}{\bibfnamefont{T.}~\bibnamefont{Nomura}} \bibnamefont{and}
  \bibinfo{author}{\bibfnamefont{H.}~\bibnamefont{Okada}},
  \bibinfo{journal}{Phys. Lett. B} \textbf{\bibinfo{volume}{783}},
  \bibinfo{pages}{381} (\bibinfo{year}{2018}), \eprint{1805.03942}.

\bibitem[{\citenamefont{Navas et~al.}(2024)}]{ParticleDataGroup:2024cfk}
\bibinfo{author}{\bibfnamefont{S.}~\bibnamefont{Navas}} \bibnamefont{et~al.}
  (\bibinfo{collaboration}{Particle Data Group}), \bibinfo{journal}{Phys. Rev.
  D} \textbf{\bibinfo{volume}{110}}, \bibinfo{pages}{030001}
  (\bibinfo{year}{2024}).

\bibitem[{\citenamefont{Kanemura et~al.}(2013)\citenamefont{Kanemura, Matsui,
  and Sugiyama}}]{Kanemura:2013qva}
\bibinfo{author}{\bibfnamefont{S.}~\bibnamefont{Kanemura}},
  \bibinfo{author}{\bibfnamefont{T.}~\bibnamefont{Matsui}}, \bibnamefont{and}
  \bibinfo{author}{\bibfnamefont{H.}~\bibnamefont{Sugiyama}},
  \bibinfo{journal}{Phys. Lett. B} \textbf{\bibinfo{volume}{727}},
  \bibinfo{pages}{151} (\bibinfo{year}{2013}), \eprint{1305.4521}.

\bibitem[{\citenamefont{Kanemura et~al.}(2016)\citenamefont{Kanemura,
  Nishiwaki, Okada, Orikasa, Park, and Watanabe}}]{Kanemura:2015bli}
\bibinfo{author}{\bibfnamefont{S.}~\bibnamefont{Kanemura}},
  \bibinfo{author}{\bibfnamefont{K.}~\bibnamefont{Nishiwaki}},
  \bibinfo{author}{\bibfnamefont{H.}~\bibnamefont{Okada}},
  \bibinfo{author}{\bibfnamefont{Y.}~\bibnamefont{Orikasa}},
  \bibinfo{author}{\bibfnamefont{S.~C.} \bibnamefont{Park}}, \bibnamefont{and}
  \bibinfo{author}{\bibfnamefont{R.}~\bibnamefont{Watanabe}},
  \bibinfo{journal}{PTEP} \textbf{\bibinfo{volume}{2016}},
  \bibinfo{pages}{123B04} (\bibinfo{year}{2016}), \eprint{1512.09048}.

\bibitem[{\citenamefont{Machacek and Vaughn}(1983)}]{Machacek:1983tz}
\bibinfo{author}{\bibfnamefont{M.~E.} \bibnamefont{Machacek}} \bibnamefont{and}
  \bibinfo{author}{\bibfnamefont{M.~T.} \bibnamefont{Vaughn}},
  \bibinfo{journal}{Nucl. Phys. B} \textbf{\bibinfo{volume}{222}},
  \bibinfo{pages}{83} (\bibinfo{year}{1983}).

\bibitem[{\citenamefont{Machacek and Vaughn}(1984)}]{Machacek:1983fi}
\bibinfo{author}{\bibfnamefont{M.~E.} \bibnamefont{Machacek}} \bibnamefont{and}
  \bibinfo{author}{\bibfnamefont{M.~T.} \bibnamefont{Vaughn}},
  \bibinfo{journal}{Nucl. Phys. B} \textbf{\bibinfo{volume}{236}},
  \bibinfo{pages}{221} (\bibinfo{year}{1984}).

\bibitem[{\citenamefont{Machacek and Vaughn}(1985)}]{Machacek:1984zw}
\bibinfo{author}{\bibfnamefont{M.~E.} \bibnamefont{Machacek}} \bibnamefont{and}
  \bibinfo{author}{\bibfnamefont{M.~T.} \bibnamefont{Vaughn}},
  \bibinfo{journal}{Nucl. Phys. B} \textbf{\bibinfo{volume}{249}},
  \bibinfo{pages}{70} (\bibinfo{year}{1985}).

\bibitem[{\citenamefont{Maki et~al.}(1962)\citenamefont{Maki, Nakagawa, and
  Sakata}}]{Maki:1962mu}
\bibinfo{author}{\bibfnamefont{Z.}~\bibnamefont{Maki}},
  \bibinfo{author}{\bibfnamefont{M.}~\bibnamefont{Nakagawa}}, \bibnamefont{and}
  \bibinfo{author}{\bibfnamefont{S.}~\bibnamefont{Sakata}},
  \bibinfo{journal}{Prog. Theor. Phys.} \textbf{\bibinfo{volume}{28}},
  \bibinfo{pages}{870} (\bibinfo{year}{1962}).

\bibitem[{\citenamefont{Esteban et~al.}(2024)\citenamefont{Esteban,
  Gonzalez-Garcia, Maltoni, Martinez-Soler, Pinheiro, and
  Schwetz}}]{Esteban:2024eli}
\bibinfo{author}{\bibfnamefont{I.}~\bibnamefont{Esteban}},
  \bibinfo{author}{\bibfnamefont{M.~C.} \bibnamefont{Gonzalez-Garcia}},
  \bibinfo{author}{\bibfnamefont{M.}~\bibnamefont{Maltoni}},
  \bibinfo{author}{\bibfnamefont{I.}~\bibnamefont{Martinez-Soler}},
  \bibinfo{author}{\bibfnamefont{J.~P.} \bibnamefont{Pinheiro}},
  \bibnamefont{and} \bibinfo{author}{\bibfnamefont{T.}~\bibnamefont{Schwetz}},
  \bibinfo{journal}{JHEP} \textbf{\bibinfo{volume}{12}}, \bibinfo{pages}{216}
  (\bibinfo{year}{2024}), \eprint{2410.05380}.

\bibitem[{\citenamefont{Casas and Ibarra}(2001)}]{Casas:2001sr}
\bibinfo{author}{\bibfnamefont{J.~A.} \bibnamefont{Casas}} \bibnamefont{and}
  \bibinfo{author}{\bibfnamefont{A.}~\bibnamefont{Ibarra}},
  \bibinfo{journal}{Nucl. Phys. B} \textbf{\bibinfo{volume}{618}},
  \bibinfo{pages}{171} (\bibinfo{year}{2001}), \eprint{hep-ph/0103065}.

\end{thebibliography}
\end{document}